\documentclass[12pt]{article}
\usepackage{graphicx}
\usepackage{psfrag}
\usepackage{epstopdf}

\begin{document}

\begin{titlepage}
\null\vspace{-62pt}

\pagestyle{empty}
\begin{center}

\vspace{1.0truein} {\Large\bf Graviton mass and cosmological constant: a toy model}

\vspace{1in}
{\large Dimitrios Metaxas} \\
\vskip .4in
{\it Department of Physics,\\
National Technical University of Athens,\\
Zografou Campus, 15780 Athens, Greece\\
metaxas@central.ntua.gr}\\

\vspace{0.5in}

\vspace{.5in}
\centerline{\bf Abstract}

\baselineskip 18pt
\end{center}

\noindent
I consider a simple model where the graviton mass and the cosmological constant depend on a scalar field with appropriate couplings and I calculate the graviton propagator and the resulting effective potential  for the scalar field in order to examine issues of stability and symmetry breaking.

\end{titlepage}

\newpage
\pagestyle{plain}
\setcounter{page}{1}
\newpage

\section{Introduction}

Infrared modifications of gravity with a small graviton mass appear in various models in cosmology and particle physics \cite{ah} and it is interesting to investigate their consistency.
Several problems appear in these models, like the Boulware-Deser instability \cite{bd}, as well as the
 van~Dam-Veltman-Zakharov (vDVZ) discontinuity that persists for arbitrarily small graviton mass \cite{vdvz1}.
It has been shown that it is possible to circumvent the latter problem in the presence of a (positive or negative) cosmological constant \cite{porr}; then, however, another problem of consistency appears: in the case of a positive cosmological constant, in particular for de~Sitter space-time, one encounters the Higuchi instability for $m_g^2 < 2 H^2$, where $m_g$ is the graviton mass and $H$ the Hubble expansion rate \cite{hig1}. It is argued that a scalar degree of freedom propagates then with the wrong sign of the kinetic term, a ghost, and several aspects of this problem have been investigated \cite{def1}. In particular, it was argued in \cite{metaxas}, that, because of the subtleties of the path integral measure in perturbative quantum gravity,
the bound for the Higuchi instability is reversed, much in the same manner that the problem of the negative conformal mode is treated in ordinary gravity.

The problem of spontaneous symmetry breaking in order to give a mass to the graviton has also been considered \cite{h1}, which is more involved than the usual process of symmetry breaking in quantum field theory, since a massive graviton has five degrees of freedom compared to the two degrees of freedom for an ordinary, massless graviton.

In order to gain some more perspective on these problems, I consider in Section~2 a simple toy model, where both the graviton mass and the cosmological constant terms depend on a scalar field, $\phi$. As far as the cosmological constant term is concerned, this is just a mass term for $\phi$. The graviton mass term is added by hand at the
quadratic perturbative level, and I calculate the effective potential induced by graviton loops, neglecting graviton self-interactions, in order to discuss the possible instabilities, in relation with the afore-mentioned
Higuchi bound. I conclude in Section~3 with some comments and discussion.

\section{Graviton propagator and effective action}

The gravitational action is augmented with a scalar field, $\phi$, with a mass term
\begin{equation}
S_1 = \int \! d^4\!x \,\sqrt{-g} \left( -\frac{R}{16 \pi G}
                            - \frac{1}{2}m^2\phi^2 \right)
\label{s1}
\end{equation}
as well as a graviton mass term
\begin{equation}
S_2 = -\int \! d^4\!x \, \frac{1}{2}\alpha \phi^2 (h_{\mu\nu}^2 - h^2)
\label{s2}
\end{equation}
at quadratic order after linearization around a flat background
\begin{equation}
g_{\mu\nu}=\eta_{\mu\nu} + \kappa\, h_{\mu\nu}
\end{equation}
where $\kappa^2 = 32 \pi G = \frac{32\pi}{M_P^2}$, $h=h_{\mu}^{\mu}$ and $\alpha$ a dimensionless coupling.

Although the action $S_1$ is invariant with respect to the gauge transformation
\begin{equation}
\delta h_{\mu\nu} = \kappa (\partial_{\mu}\epsilon_{\nu} + \partial_{\nu}\epsilon_{\mu}),
\label{gaugetr}
\end{equation}
corresponding to a massless graviton with two degrees of freedom,
the addition of $S_2$ breaks this invariance and describes a massive graviton with five degrees of freedom
and mass $m_g^2 = \alpha \phi^2$.

If the graviton mass term is considered as an interaction, as is done here, one still needs to gauge fix the action $S_1$ with the Stueckelberg procedure, where functional integration over $\epsilon_{\mu}$ is done, in order to introduce a gauge fixing term. This will generate additional terms coming from the non-invariant $S_2$ which I will describe later.

Adding the gauge fixing term, $(\partial^{\mu} h_{\mu\nu} - \frac{1}{2}\partial_{\nu} h)^2$, the terms
of $S_1$ that are quadratic in $h_{\mu\nu}$ become
\begin{equation}
L_1= -\frac{1}{2} h_{\mu\nu} \, \partial^2 \, K^{\mu\nu,\beta\rho} h_{\beta\rho}
      -\frac{1}{2} \lambda\phi^2 \left(h_{\mu\nu}h^{\mu\nu}-\frac{1}{2} h^2 \right)
\label{l1}
\end{equation}
where $\lambda = \frac{8 \pi m^2}{M_P^2}$ and
\begin{equation}
K^{\mu\nu,\beta\rho} = \frac{1}{2}(\eta^{\mu\beta}\eta^{\nu\rho}+\eta^{\mu\rho}\eta^{\nu\beta})
                       -\frac{1}{2} \eta^{\mu\nu} \eta^{\beta\rho}
\end{equation}
or, more compactly,
\begin{equation}
K = I -\frac{1}{2} gg.
\end{equation}
Since $K^2 = I$, the inversion of the first term in $L_1$ in order to get the bare graviton propagator is straightforward,
\begin{equation}
G=\frac{i}{k^2} (I -\frac{1}{2} gg),
\end{equation}
as is the inclusion of the ``cosmological constant'' interaction term, $-i\lambda\phi^2(I-\frac{1}{2}gg)$,
to get
\begin{equation}
G_0=\frac{i}{k^2 - \lambda\phi^2} (I - \frac{1}{2} gg)
\end{equation}
(where $k$ is the graviton momentum).

The resummation of the interaction corresponding to the graviton mass, $-i\alpha\phi^2 (I-gg)$, can also be done,
yielding the final dressed graviton propagator
\begin{equation}
\tilde{G} = \frac{i}{k^2 -\lambda\phi^2 -\alpha\phi^2}(I-\frac{1}{4}gg)
           -\frac{i}{k^2 -\lambda\phi^2 -3\alpha\phi^2}(\frac{1}{4}gg).
\label{p1}
\end{equation}
It is interesting to see that, even in the limit of $\lambda=0$, and writing $m_g^2=\alpha\phi^2$ for the graviton mass, one gets for the massive graviton propagator
\begin{equation}
G_{m_g} = \frac{i}{k^2 - m_g^2}(I-\frac{1}{4}gg)
           -\frac{i}{k^2 -3 m_g^2}(\frac{1}{4}gg),
\label{pm}
\end{equation}
a different expression than what is normally used, that does not suffer from the vDVZ discontinuity. Of course, it is well known that, when this is absent, there are problems of ghost-like states \cite{bd, peter}, and indeed, one can see that in (\ref{pm}) there are five degrees of freedom corresponding to a massive graviton, plus two longitudinal degrees of freedom, one normal and one ghost-like, with poles at different positions that do not cancel:
the first term in (\ref{pm}) can be written in terms of the massive graviton polarization tensor, $I-\frac{2}{3}gg$,
plus a longitudinal mode with normal sign as opposed to the second, ghost-like longitudinal term. In the infrared limit, the sum of all these modes gives the usual polarization tensor, $I-\frac{1}{2}gg$, so that the theory does not suffer from the vDVZ discontinuity. The two longitudinal modes obviously do not cancel for arbitrary $k$, leading to the usual ghost problems which
would create consistency problems for the full theory. At this order, however, one can calculate an effective potential for the scalar field using the dressed propagator, $\tilde{G}$ from (\ref{p1}), to get

\begin{equation}
U(\phi)= \frac{1}{2}m^2\phi^2  +
        \frac{(\lambda^2-3 \alpha^2)}{32\pi^2} \phi^4 \left( \ln{\frac{\phi^2}{m^2}}
                                                                   -\frac{25}{6}\right).
\label{ueff}
\end{equation}
Since there is already a scale, $m$, in our toy model, 
I have used the renormalization conditions 
$U''(0)=m^2$ and $U^{(4)}(m)=0$.
Now, one can vary the parameters, $\alpha$ and $\lambda=8\pi m^2/M_P^2$, and examine the resulting
effective potential, along with the dimensionless ratio,
\begin{equation}
\xi(\phi)=\frac{16 \pi \,U(\phi)}{3 \alpha \,\phi^2 \, M_P^2},
\end{equation}
which is 
the generalization of the ratio $2 H^2/m_g^2$, that is related to the Higuchi bound.
In Figs.~1-5, some examples of $U(\phi)$ and $\xi(\phi)$ are shown as functions of $\phi$
for various values of the parameters
(the dimensionful quantities are in appropriate powers of $M_P$).

Before discussing the results I should comment on the fact that the scalar field was considered here without a kinetic term, which is, anyway,
expected to be generated by gravitational corrections. In fact, a similar calculation can be done with a kinetic term along the lines of \cite{smolin},
where a general potential term was also used. Since we are interested here in the interplay between the graviton mass and the scalar mass and the resulting instabilities,
this is not necessary; the conclusions are not expected to change in a qualitative manner. Without a scalar kinetic term one can also consider
the scalar field as a Lagrange multiplier term, and the action as a nonlinear generalization of massive gravity which may develop instabilities for some range of
parameters.

Also, when integrating over the gauge parameter, $\epsilon_{\mu}$, in order to introduce the gauge fixing term, one gets additional terms because of the non-invariance of $S_2$:
\begin{equation}
\delta L = \alpha\kappa^2\phi^2\left(\epsilon_{\mu}\partial^2 \epsilon^{\mu} +(\partial\epsilon)^2
           +2 h (\partial\cdot\epsilon)\right)
\end{equation}
(this expression has been derived for constant $\phi$ and the gauge condition $\partial^{\mu}h_{\mu\nu}=\frac{1}{2}\partial_{\nu}h$ has been used).
The integration over $\epsilon_{\mu}$ can now be done using an additional gauge fixing for the abelian gauge invariance of the last expression, leading to an additional $\phi$-dependent term. For values of the parameters such that $\alpha\phi^2$ is much smaller than $ M_P^2$ the variation of this term can be neglected (and when the gauge condition $\partial\cdot\epsilon=0$ is used these terms will decouple from the gravitons to this order).

Keeping in mind the previous approximations, as well as the fact that the quantities $m^2$ and $U(\phi)$ should be of a smaller order of magnitude
than $M_P^2$ and $M_P^4$ respectively, in order for the gravitational corrections to be negligible \cite{smolin},
it is generally suggestive from the examination of Figs.1-5, that the inverse Higuchi bound is favored \cite{metaxas}.

Although the areas where $U(\phi)$ is negative are not obvious to interpret,
since the Higuchi bound concerns de~Sitter space-time background,
regions of stability of the perturbative effective potential generally correspond to values of $\xi(\phi)$ larger than unity.
It is not clear how far away from the origin of the effective potential graph our approximations are valid,
and if, for example, a symmetry breaking minimum such as the one of Fig.~2, appears at values that are consistent
perturbatively. Generally, more complete calculations, including non-renormalizable terms \cite{smolin}, as well as physical considerations,
purport that this is not the case.

 However,
 a situation such as the one shown  in Fig.~5,
seems more promising,
 with a secondary, metastable minimum appearing,
at a point where the inverse Higuchi bound is almost saturated,
and for smaller values of $\phi$ and $m$ than those of Fig.~2, corresponding to a theory with a massive graviton in de~Sitter space-time
that can decay to ordinary flat space-time with a massles graviton. This is a configuration that is, in principle, 
physically possible, and may be of interest in a cosmological context.

\section{Comments}

In this work I used a simple model where the graviton mass (as well as the cosmological constant) are obtained through the couplings of a scalar field, $\phi$ with the graviton field. I obtained the effective potential for $\phi$ and examined its stability with respect to the variations of the parameters of the model and in relation with the Higuchi bound for stability 
of a massive graviton in de~Sitter space-time; for most cases of the parameters the inverse Higuchi bound derived in \cite{metaxas} is favored.
An interesting case, for particular values of the parameters, was also found, with a metastable vacuum for a non-zero value of $\phi$ that corresponds to 
a massive graviton in de~Sitter space-time that can decay to a massless graviton in flat space-time.

As far as extensions of this model are concerned, it would be possible to add a small cosmological term to the tree level action and consider corrections similarly to the work of \cite{vilenkin1}; similar conclusions are expected provided the corrections are within the limits of the approximations.
It would also be interesting to see if the conditions for an inflationary model can be satisfied for some values of the parameters, or after modifying the present model using two different scalar fields, similarly to a hybrid model, and to check the parameter space of the full effective action.

\vspace{0.4in}

\centerline{\bf Acknowledgements}
 \noindent 
This work was completed in
the National Technical University of Athens. I am grateful to the
people of the Physics Department for their support.

\newpage

\begin{figure}
\centering
\includegraphics[width=90mm]{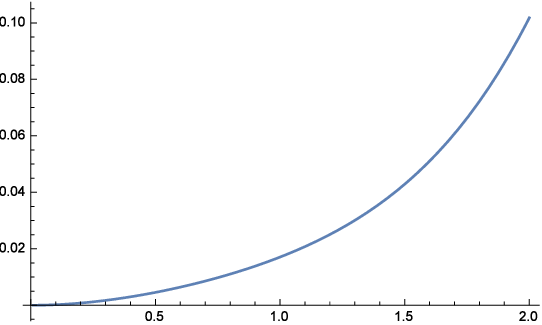}
\includegraphics[width=90mm]{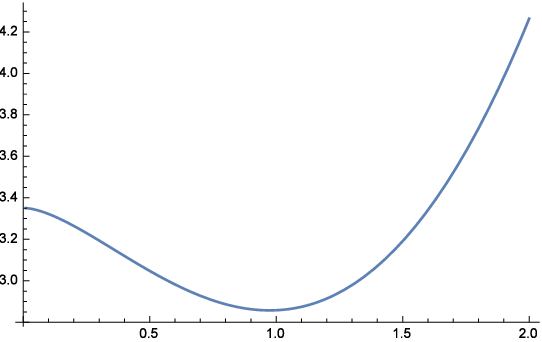}
\caption{The effective potential, $U(\phi)$ (top), and the dimensionless quantity, $\xi(\phi)$ (bottom) as functions of $\phi$, in appropriate units of $M_P$,
for values of the parameters $m=0.2$, $a=0.1$.}
\end{figure}

\begin{figure}
\centering
\includegraphics[width=90mm]{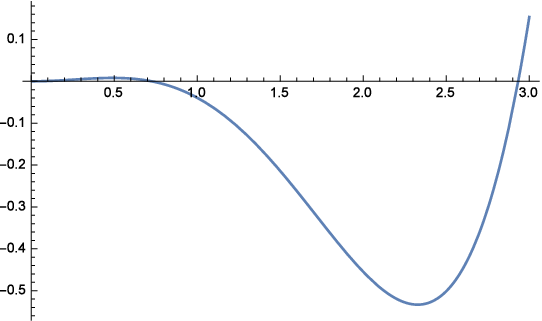}
\includegraphics[width=90mm]{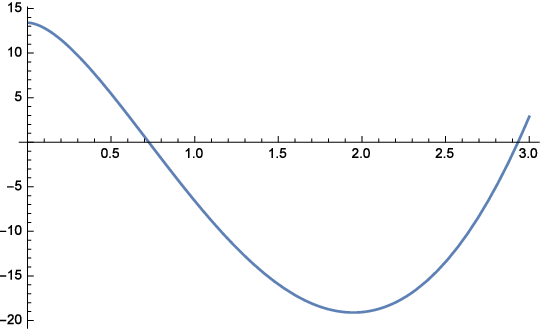}
\caption{Same as Fig.~1 for values $m=0.4$, $a=0.1$. }
\end{figure}

\begin{figure}
\centering
\includegraphics[width=90mm]{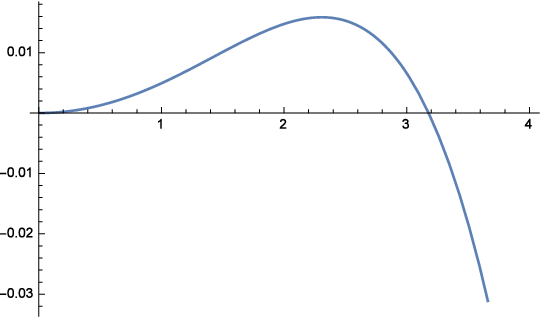}
\includegraphics[width=90mm]{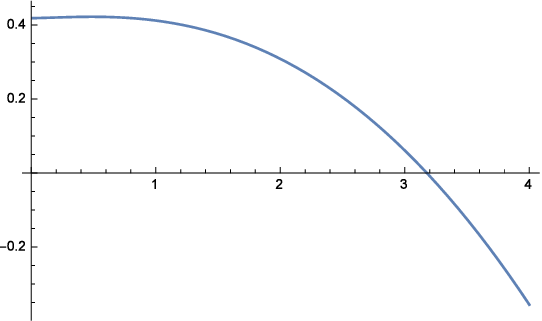}
\caption{Same as Fig.~1 for values $m=0.1$, $a=0.2$.}
\end{figure}

\begin{figure}
\centering
\includegraphics[width=90mm]{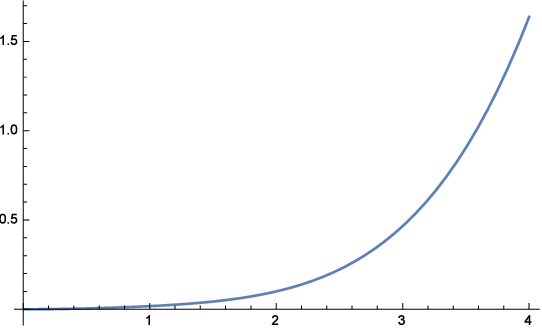}
\includegraphics[width=90mm]{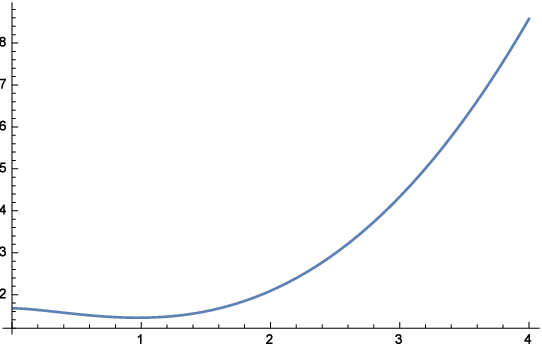}
\caption{Same as Fig.~1 for values $m=0.2$, $a=0.2$.}
\end{figure}

\begin{figure}
\centering
\includegraphics[width=90mm]{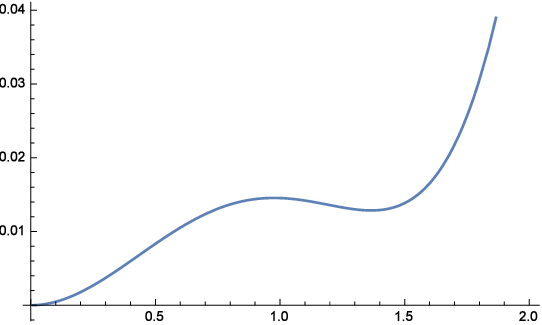}
\includegraphics[width=90mm]{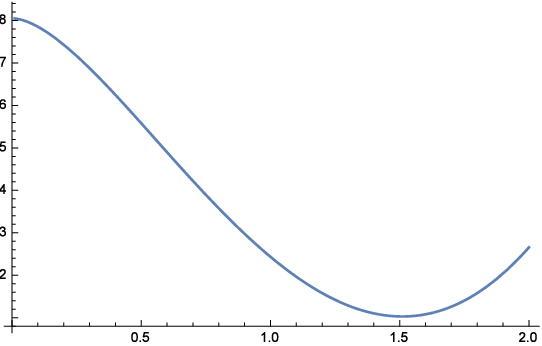}
\caption{Same as Fig.~1 for values $m=0.31$, $a=0.1$.}
\end{figure}

\end{document}